\documentclass{llncs}
\usepackage{latexsym}
\usepackage{amsmath}
\usepackage{amssymb}
\usepackage{color}
\usepackage{graphics}
\usepackage{epsfig}

\begin{document}

\title{Design of a Distributed Reachability Algorithm for Analysis of Linear Hybrid Automata}
\author{Sumit Kumar Jha }
\institute {Computer Science Department, Carnegie Mellon University,
Pittsburgh PA 15213}

%%\author{Edmund M. Clarke  \inst{1},  Sumit K. Jha \inst{1}, Bruce H. Krogh \inst{2} ,  Prasad Sistla \inst{3}, Jim Weimer \inst{2} }
%%\institute{ Computer Science Department, Carnegie Mellon University\\
%%5000 Forbes Avenue, Pittsburgh, PA 15213,
%%USA\\emc+ , jha+ @cs.cmu.edu \\
%%\and ECE Department, Carnegie Mellon University\\
%%5000 Forbes Avenue,
%%Pittsburgh, PA15213, USA\\krogh@ece.cmu.edu \\
%%\and Computer Science Department, University of Illinois, Chicago, IL, USA\\
%%sistla@cs.uic.edu\\%

\maketitle

\abstract{ This paper presents the design of a novel distributed algorithm d-IRA for the reachability analysis of linear hybrid automata. Recent work on \textit{iterative relaxation abstraction} (IRA) is leveraged to distribute the computational problem among multiple computational nodes in a non-redundant manner by performing careful infeasibility analysis of linear programs corresponding to spurious counterexamples. The  d-IRA algorithm  is resistant to failure of multiple computational nodes. The experimental results provide promising evidence for the possible successful application of this technique.}

\section{Introduction}

The verification of hybrid systems is a computationally expensive procedure and often does not succeed except for systems with a few continuous variables. Linear hybrid automata are an important class of hybrid systems which can approximate nonlinear hybrid systems in an asymptotically complete fashion~\cite{ho}. We extend earlier work~\cite{jha2007} on applying counterexample guided abstraction refinement (CEGAR) algorithms to the analysis of linear hybrid automata and present a distributed algorithm for their reachability analysis.

We believe that a distributed analysis engine for hybrid automata will be an important achievement in the area of analyzing hybrid systems. There are two important developments that motivate our research in this direction:

\begin{itemize}

\item While the computational power on a single core processor had  traditionally been growing exponentially, recent trends ~\ cite [intel, amd] by microprocessor manufacturers have clearly indicated that this exponential growth is no longer feasible. This requires that computational systems
adapt to the hardware systems that are going to be available in the future.  While efficient compiler and hardware techniques were successful at finding parallelism in programs running on a single core processor, the advent of multiple processors on a single chip puts the burden of finding parallelism more on the designer of the algorithm and the architect of the software system. In particular, software systems and the algorithms they implement must \emph{explicitly} provide opportunities for multiple cores to be in use simultaneously. \\

\item Borrowing techniques from the domain of linear programming and ideas from the realm of practically applied software verification, we have recently proposed an \textit{iterative relaxation abstraction} technique which exploits the structure of a linear hybrid automata while trying to solve its reachability analysis problem. The IRA algorithm is based on solving several smaller sub-problems one after another instead of solving the original problem at once. Further reflection shows that several subproblems being constructed can be built and solved in a distributed manner, with a relatively small amount of book keeping and further algorithmic analysis.\\

\end{itemize}

This paper makes several novel contributions to the development of practical algorithms for the analysis of linear hybrid automata:\\

1. We present the first distributed algorithm for the analysis of linear hybrid automata. Our algorithm is tolerant of multiple failures in the distributed computational nodes.

2.  We present new theoretical results establishing a partial-order among counterexamples and relaxations of linear hybrid automata. Using these results, we find several counterexamples not related by the partial order and build relaxations to refute each of them in a distributed manner.

3. We show that the distributed system has a small global state which needs to be preserved in case of failure of the distributed system; we also identify the potential to backup this global state without slowing down the distributed computation.

\section{Related Work}

The reachability analysis algorithms for linear hybrid automata have been studied in~\cite{ho}. . These algorithms continue to be the driving horse for PHAVer~\cite{DBLP:conf/hybrid/Frehse05}, IRA~\cite{jha2007} and our current techniques too. We have built upon these core algorithms and did not intend to replace them.

Our current work is inspired by our development and analysis of several LHA examples using the IRA algorithm, which is now re-implemented within PHAVer.
The IRA algorithm is essentially a CEGAR based technique for analyzing Linear Hybrid Automata (LHA). IRA introduces the idea of constructing multiple relaxations of an LHA and proves the reachability property over the original LHA using these relaxations. Each relaxation of the LHA is an over-approximate abstraction for the LHA but the relaxed LHA often involves relatively fewer number of continuous variables and is, hence, more amenable to analysis. First, IRA~\cite{jha2007} constructs a relaxation $H_i$ of the LHA $H$. It then queries the underlying LHA reachability engine like PHAVer~\cite{DBLP:conf/hybrid/Frehse05} and
builds an over-approximate discrete abstraction ( a regular language FSM ) $A_i$ for the relaxed hybrid automata $H_i$. If the language of the over-approximate discrete abstraction contains no counterexamples, the bad state is not reachable in the original LHA either and the algorithm terminates. Otherwise, IRA picks up a
counterexample from the discrete abstraction and constructs a linear program to check for its validity in the high dimensional original linear hybrid automata. If the linear program is satisfiable, the counterexample is valid~\cite{jha2006} and hence, the IRA algorithm  reports that the bad sate is reachable and stops. Otherwise the linear program is infeasible, a small subset of variables is identified using infeasibility analysis and a new relaxed hybrid automata $H_{i+1}$ is constructed using these variables. This standard algorithm is discussed in \cite{jha2007}.

Our experience with the development and analysis of examples using our tool IRA showed that several expensive computations performed by the IRA reachability engine can be performed independently in a distributed manner.

\section{The Distributed Algorithm (d-IRA) }

\begin{figure}[htbp]
\input{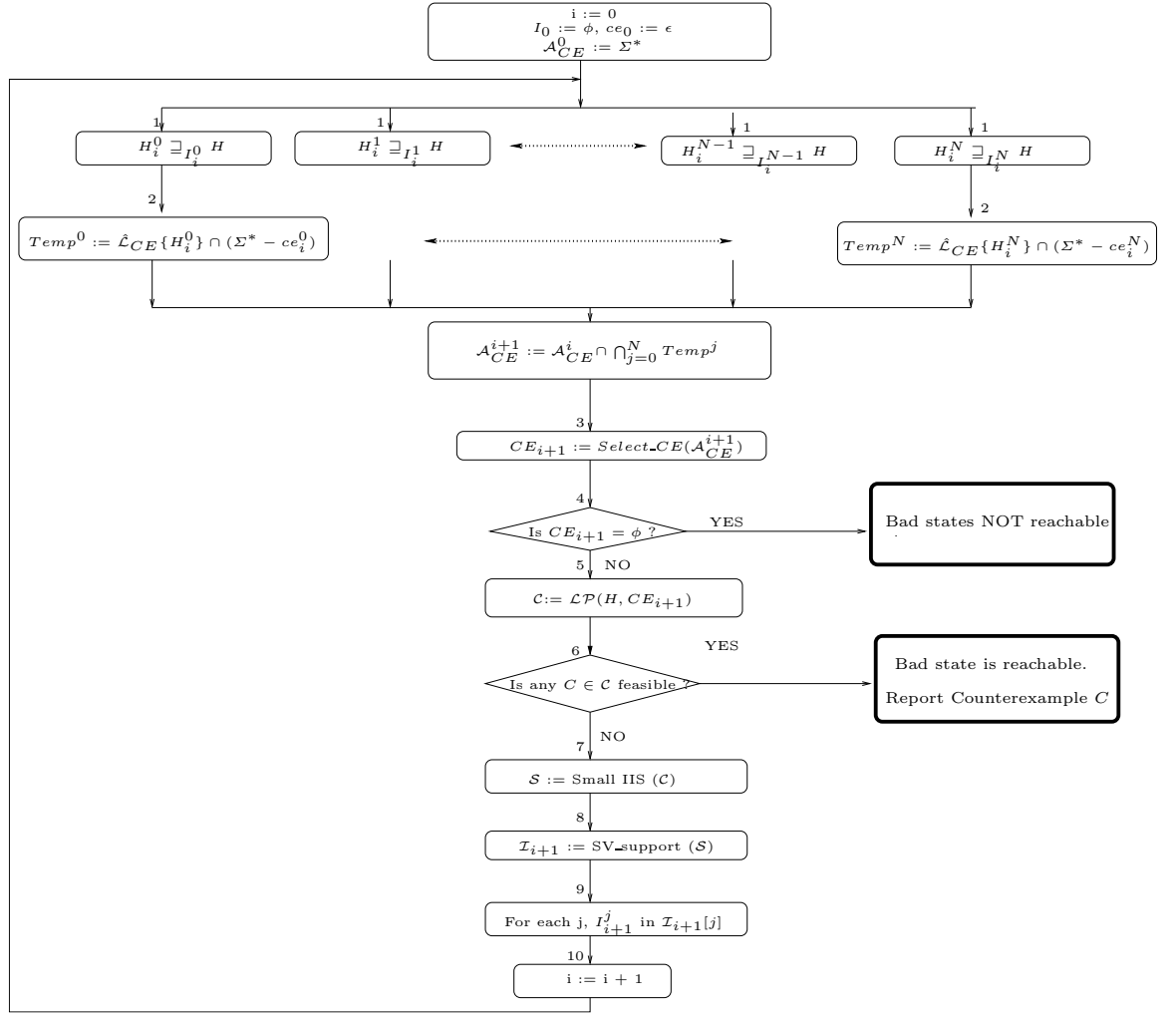}
\caption{The d-IRA procedure: Distributed Iterative Relaxation
Abstraction. Note that the expensive calls to the underlying hybrid
automata reachability engine occurs in parallel. } \label{algorithm}
\end{figure}

In this paper, we present a distributed version of the IRA algorithm. The flow of the algorithm is sketched in Fig.~\ref{algorithm}. The distributed algorithm assumes one master computation node and (N-1) other computational (slave) nodes. Initially, the master node initialises a counter $i$ to zero, chooses the empty set as an initial set of variables $\mathcal{I}_0$ and learns the deterministic finite automata corresponding to $\Sigma^*$ as the initial discrete over-approximate \emph{global abstraction} of the language of the LHA $H$.

1. During the $i^th$ iteration, the $j^{th}$ computational node constructs its own relaxation $H_i^j$ of the linear hybrid automata $H$ using the set of variables $I_i^j$. It should be noted that the process of constructing relaxations may also be computationally very intensive and may involve invoking the Fourier-Motzkin elimination routine.

2. Each computational node then constructs a discrete abstraction $Temp^j$ corresponding to the relaxed linear hybrid automata $H_i^j$. This step involves making calls to the underlying reachability engine like PHAVer~\cite{DBLP:conf/hybrid/Frehse05}.

Both the above steps are identical to the corresponding steps in the IRA algorithm~\cite{jha2007} and are not discussed here for brevity.

3. Each computational node sends the discrete abstraction $Temp^j$ which it learned from  the relaxed linear hybrid automata $H_i^j$ to the master computational node. This is the only step at which there is communication from the slave nodes to the master node during the d-IRA algorithm.

4. The master node updates the discrete \emph{global  abstraction} $A_{CE}^{i+1}$ by taking the intersection of the previous  discrete \emph{global  abstraction} $A_{CE}^i$ with all the newly learned discrete abstractions $Temp^j$.

5. Then, the master picks a set ${CE}$ of $N$ non-redundant counterexamples from the newly built discrete \emph{global abstraction} $A_{CE}^{i+1}$. This is an algorithmically interesting step and is detailed in Section~\ref{partial-order}.

6. The master node checks if the set of counterexamples ${CE}$ is empty. If  $A_{CE}^{i+1}$ has no counterexamples, then no bad states are reachable in the system~\cite{jha2007} and hence, it is declared to be safe.

7. The master computational node forms a set of linear programs $\mathcal{C}$, where each linear program corresponds to one of the counterexamples in $CE_{i+1}$. This step is similar to the corresponding step in the IRA algorithm ~\cite{jha2007} and is discussed in depth in ~\cite{jha2006}.

8. The master node checks if any of the linear programs in $\mathcal{C}$ is feasible. In any of them, say $C$, is feasible, we stop and report that the bad state is reachable~\cite{jha2006}. We also report the counterexample corresponding to the linear program $C$.

9. If none of the linear program are feasible, the master node finds the \emph{irreducible infeasible subsets} for each of the linear programs.

10. The master node uses the basis of the IIS as the choice for the next set of variables $\mathcal{I}_{i+1}$ which will be used to construct the relaxations. The master node communicates the set $I_{i+1}^j$ to the $j^{th}$ client. This is the only step in the d-IRA algorithm during which the master sends messages to the client nodes.

10. The master then increment the counter $i$ and goes back to the distributed computation at Step 2.

\section{A Partial Order for Counterexamples and Relaxations}
\label{partial-order}

In order to make the distributed computation effective, it is essential that the various computational nodes do not solve equivalent reachability sub-problems. In particular, we want to make sure that the relaxed linear hybrid automata for the $i^{th}$ iteration $H_i^j$ and $H_i^k$ are different\footnote{We note that the property of IRA that no two relaxations across the iterations are identical still holds and with the same proof}. We achieve this goal by making a suitable choice of counterexamples from the global abstraction $A_{CE}^{i+1}$. Before we present our algorithmic methods, we define some related notions. Our definitions of linear hybrid automata, relaxations and counterexamples  are identical to those in literature ~\cite{ho,jha2007} and we do not repeat them here for sake of brevity. Given a path $\rho$ in a linear hybrid automata $H$, we can derive a set of corresponding linear constraints $Constraints(H,\rho)$ which is feasible if and only if the path is feasible. This construction\cite{jha2006,jha2007} is omitted here.

\begin{definition}
\emph{Minimal Explanation for Infeasible Counterexamples} : Given a counterexample path $\rho$ which is infeasible in a linear hybrid automata H but feasible in a relaxation $H'$ of $H$, (i.e. $H' \sqsubseteq H$), a set of linear constraints $IIS(\rho)$ is said to be an IIS for $\rho$ if and only if:

\begin{itemize}
\item $IIS(\rho) \subseteq Constraints(H,\rho)$
\item $IIS(\rho)$ is not feasible.
\item for any set S s.t. $S \subset IIS(\rho)$, S is feasible.
\end{itemize}

The special basis $Var$ of the IIS of $\rho$  is called a \emph{minimal explanation for the infeasible counterexample} and we write it as $Var(\rho, IIS(\rho) )$.

\end{definition}

In the following, we assume that there exists a function $\mathcal{IIS}$ which maps each counterexample to a unique IIS.

\begin{definition}
\emph{Dominance of Counterexamples} : A counterexample $ce$ is said to dominate a counterexample $ce'$  if and only if  $Var(ce,\mathcal{IIS}(ce)) \subseteq Var(ce',\mathcal{IIS}(ce'))$. We write $ce
{\succeq}  ce'$.
\end{definition}

We now define the notion of equivalent counterexamples and show that dominance relation among counterexamples forms a partial order.

\begin{definition}
Two counterexamples $ce$ and $ce'$ are said to be equivalent if and only if $Var(ce,\mathcal{IIS}(ce))$ $=$ $Var(ce',\mathcal{IIS}(ce'))$. Then, we say $ce \approx ce'$.

\end{definition}

\begin{theorem}
The dominance relation $\succeq$ among counterexamples is a partial order relation.
\end{theorem}
\begin{proof}
We prove that $\succeq$ is reflexive, antisymmetric and transitive:

\emph{Reflexivity}: For every counterexample, $Var(ce,\mathcal{IIS}(ce))$ $\subseteq$ $Var(ce,\mathcal{IIS}(ce))$; hence, $ce \succeq ce$.

\emph{Antisymmetry}: Suppose $ce \succeq ce'$ and  $ce' \succeq ce$. Then, $Var(ce,\mathcal{IIS}(ce))$ $\subseteq$ $Var(ce',\mathcal{IIS}(ce'))$, and also, $Var(ce',\mathcal{IIS}(ce'))$ $\subseteq$ $Var(ce,\mathcal{IIS}(ce))$. Thus, $Var(ce,\mathcal{IIS}(ce))$ $=$ $Var(ce',\mathcal{IIS}(ce'))$.Hence, $ce \approx ce'$.

\emph{Transitivity}: Suppose $ce \succeq ce'$ and  $ce' \succeq ce''$, then   $Var(ce,\mathcal{IIS}(ce))$ $\subseteq$ $Var(ce',\mathcal{IIS}(ce'))$ and $Var(ce',\mathcal{IIS}(ce'))$ $\subseteq$ $Var(ce'',\mathcal{IIS}(ce''))$. Thus, $Var(ce,\mathcal{IIS}(ce))$ $\subseteq$ $Var(ce'',\mathcal{IIS}(ce''))$. Hence, $ce \succeq ce''$.

\end{proof}

\begin{theorem}
The relaxations $\{H_i\}$ of $H$  form a partial order.
\end{theorem}
\begin{proof}
We prove that the relaxation relation $\sqsubseteq$ among $H_i$ is a partial order. \emph{Reflexivity}: For every relaxed hybrid automata, $H_i$ $\sqsubseteq$ $H_i$. \emph{Antisymmetry}: Suppose $H_i \sqsubseteq H_j$ and  $H_j \sqsubseteq H_i$. Then, $H_i = H_j$.
\emph{Transitivity}: Suppose $H_i \sqsubseteq H_j$ and  $H_j \sqsubseteq H_k$, then   $H_i \sqsubseteq H_k$.

\end{proof}

\begin{theorem}
Let $H_{ce}$ be the relaxation of $H$ w.r.t. $Var(ce,\mathcal{IIS}(ce))$  and $H_{ce'}$ be the relaxation of $H$ w.r.t. $Var(ce',\mathcal{IIS}(ce'))$. If the counterexample $ce$ dominates the counterexample $ce'$ i.e. $ce \succeq ce'$, then $H_{ce}$ is a relaxation of $H_{ce'}$ i.e. $H_{ce} \sqsubseteq H_{ce'}$.

\end{theorem}

\begin{proof}
Since $ce \succeq ce'$, $Var(ce,\mathcal{IIS}(ce))$ $\subseteq$ $Var(ce',\mathcal{IIS}(ce'))$. Thus, $H_{ce} \sqsubseteq_{Var(ce,\mathcal{IIS}(ce)) \setminus Var(ce',\mathcal{IIS}(ce')) }  H_{ce'} \sqsubseteq_{Var(ce',\mathcal{IIS}(ce'))} H$.

\end{proof}

The algorithm for selecting N counterexamples is based on the above results.\\

\framebox{
\begin{minipage}{10cm}
Algorithm \emph{Select\_CE}

Input: Global Abstraction Automata $A_{CE}^i$, LHA $H$, a timer TIMEOUT.

Output: N counterexamples: ${CE} = \{ ce_1, \dots ce_N \}$\\

1. Initialize $CE$ to be the empty set.

2. Pick a set of m $(>N)$ distinct counterexamples $C = \{ce_1, ce_2 \dots ce_m \}$ from  $A_{CE}^i$.

3. Build a set of linear programs $\{lp_1, lp_2 \dots lp_m \}$ corresponding to each of $\{ce_1, ce_2 \dots ce_m \}$

4. For each (infeasible) linear program $lp_i$, obtain an IIS and remember it as $\mathcal{IIS}(lp_i)$

5. For each counterexample $ce_i \in C $,

\hspace{0.3cm} a. Check whether there exists a counterexample $ce_j \in C $ such that $ce_j \succeq ce_i$ ($i \neq j$).

\hspace{0.5cm} b. If no such counterexample $ce_j$ exists, add $ce_i$ to $CE$.

\hspace{0.5cm} c. Remove $ce_i$ from $C$.

6. If ( $|CE| < N$ $and$ $! TIMEOUT$ ) , $m = m \times 2$ ; goto step 2.

7. Assert ($|CE| \geq N$ $or$ $TIMEOUT$); RETURN the first N members of CE as a set.

\end{minipage}
}

\section{Properties and Extensions of d-IRA}

%\subsubsection{Properties of d-IRA}
\begin{theorem}
The d-IRA algorithm is resistant to failures and restarts of all slave nodes.
\end{theorem}
\begin{proof}
If the $i^{th}$ slave node fails during the $j^{th}$ iteration, then the d-IRA algorithm can still proceed by making the assumption that $L(Temp_i) = \Sigma^*$. When the $i^{th}$ node has recovered, it can continue to participate from the next iteration.
\end{proof}

The resistance to failures of slave computational nodes is possible because the slave nodes do not store any global state information during the distributed computation and the overall distributed reachability computation itself does not depend critically on one or more slave nodes. It is also to be noted that the communication bandwidth is bounded by the sum of the sizes of the discrete abstractions and the variable sets at each stage.

\subsubsection{Tolerance to Failure of Master Computation Node}

The d-IRA algorithm depends critically on the master computational node and its failure would prematurely end the distributed computation. While master nodes could be chosen to be very reliable, the algorithm can also be adapted to handle unreliable master nodes.

It is to be observed that the d-IRA algorithm spends most of its time performing relaxations and reachability computations during which the communication infrastructure would remain idle. Further, the current state of the distributed computation is really captured completely by the global abstraction $A_{CE}^i$ after the $i^{th}$ iteration. It is hence desirable to communicate the global abstraction to either a group of \emph{shadow masters} or to the slave machines themselves during periods of low communication activity. In such a scenario, the failure of the master node would only require that an old copy of the   global abstraction $A_{CE}^i$ be obtained from one of the shadow masters or the slave machines. Then, the distributed computation would restart without wasting the computations already completed in the first $i$ iterations.

\begin{theorem}
The modified $d-IRA$ algorithm is resistant to failures and restarts of the master node.
\end{theorem}

\section{Experimental Results and Conclusion}

We implemented a version of our distributed algorithm using the IRA infrastructure. We ran our experiments on a four processor 64-bit  AMD Opteron(tm) 844 SMP
machine running Red Hat Linux version 2.6.19.1-001-K8. We only implemented a parallel version of the relaxation step to test the validity of these ideas. We found up to a 3.41-X increase in performance on our four processor machine with this implementation on a set of parameterized adaptive cruise control examples~\cite{jha2007}.

\begin{table*}
\caption{Distributed IRA vs IRA}
\label{tab:perf1} \centering
\renewcommand{\arraystretch}{1.2}
\setlength\tabcolsep{3pt}
\begin{tabular}[t]{@{}lcccp{0cm}rr@{}}
Example & $\sharp$-Variables & Time for d-IRA [s] & Time for IRA [s] \\
\noalign{\smallskip} \hline \noalign{\smallskip}
ACC-4 & 4 & 11 & 15 \\
ACC-8 & 8 & 100 & 192 \\
ACC-16 & 16 & 1057 & 3839   \\
ACC-19 & 19 & 2438 & 9752 \\
\hline
\end{tabular}
\end{table*}

A case for the architecture of a distributed hybrid
systems model checker has been made in this paper
which uses the partial order relation among multiple
counterexamples obtained during the Iterative
Relaxation Abstraction procedure for the generation of non-redundant distributed sub-problems.

\bibliography{distributed-hybrid}

\end{document}